\documentclass[11pt]{article}\topmargin=-0.5cm\hfuzz=10pt  \oddsidemargin=-0.1cm \textheight 222mm\textwidth=16.3cm
\newcommand{\comm}[1]{}

\usepackage{amssymb,epsfig}
\usepackage{amsthm}
\usepackage[numbers]{natbib}
\def\citet{\cite}
\usepackage{etoolbox}
 \usepackage{color}

\newtheorem{theorem}{Theorem}
\newtheorem{definition}{Definition}
\newtheorem{remark}{Remark}

\def\e{\varepsilon}

\def\defi{\stackrel{{\scriptscriptstyle \Delta}}{=}}

\def\d{\delta}
\def\o{\omega}

\def\Y{{\cal Y}}
\def\F{{\cal F}}
\def\w{\widehat}
\def\Re{{\,\rm Re\,}}
\def\Im{{\,\rm Im\,}}
\def\Ind{\mathbb{I}}

\def\R{{\bf R}}

\def\S{{\bf S}}
\def\Z{{\cal Z}}

\def\C{{\bf C}}

\def\ww{\widetilde}
\def\X{{\cal X}}

\def\oo{\bar}

\newcommand{\be}{\begin{equation}}
\newcommand{\ee}{\end{equation}}
\newcommand{\bd}{\begin{displaymath}}
\newcommand{\ed}{\end{displaymath}}
\newcommand{\ba}{\begin{array}{ll}}
\newcommand{\ea}{\end{array}}
\newcommand{\baa}{\begin{eqnarray}}
\newcommand{\eaa}{\end{eqnarray}}
\newcommand{\baaa}{\begin{eqnarray*}}
\newcommand{\eaaa}{\end{eqnarray*}}

\def\oo{\bar}

\def\y{\mu}

\def\Re{{\rm Re\,}}

\def\ew{\left(e^{i\o}\right)}
\def\T{{\mathbb{T}}}
\def\ZZ{{\mathbb{Z}}}

\def\TT{{\cal T}}

\def\F{{\cal F}}

\def\supp{{\rm supp}\,}
\date{Submitted December 4 2017; revised October 3 2018}
\title{On data recovery with restraints on the spectrum range and the process range}
\author{
Nikolai Dokuchaev \comm{School of Electrical Engineering, Computing and Mathematical Sciences, Curtin
University, GPO Box U1987, Perth, 6845 Western Australia.Email: N.Dokuchaev@curtin.edu.au}}
\begin{document}
\def\break{}%
\def\brea{}
\def\breakk{}
\vspace{1cm}
\maketitle
\let\thefootnote\relax\footnote{ The author is with
School of Electrical Engineering, Computing and Mathematical Sciences
University, {  GPO Box U1987, Perth, 6845 Western Australia.
 \comm{ and National Research University ITMO, 197101 Russia. } }
 }
 \begin{abstract}  The paper considers recovery of signals from incomplete observations and a problem of determination  of the allowed quantity of missed observations, i.e. the problem  of determination  of the size of the uniqueness sets for a given data recovery procedures.  The  paper suggests a way to bypass  solution of this uniqueness problem  via   imposing restrictions
      investigates possibility of data recovery
   for classes of
 finite sequences under a special discretization of the process range.
 It is shown that these  sequences can be
dense in the space of all sequences and that the
uniqueness sets for them can be  singletons. Some robustness with respect to rounding of input data can be achieved via including additional  observations.

 \par
{\bf Key words}: data recovery, data compression, discrete Fourier transform,
spectrum range discretization,  process range discretization, Diophantine equations.
\end{abstract}

\vspace{-0.5cm}
\section{Introduction}
The paper investigates possibility of recovery of finite sequences from partial observations
 in the setting with insufficient statistics
where the probability distributions are unknown. In other words, this setting is oriented on the
sequences being considered as sole sequences rather  than members of an ensemble; this feature is typical
in  signal processing for speech, images, human activity analysis, traffic  control, etc.

The data recovery problem was studied intensively  in different settings exploring different restrictions on classes of underlying processes classes.  Usually,  recoverability   is associated with sparsity or certain  restrictions on the spectrum support
such as  bandlimitiness or the presence of spectrum gaps; see  e.g.
\cite{CT06,CR1,D16a,OU} and references therein.
 The classical
Nyquist-Shannon sampling theorem  establishes that a band-limited continuous time function can be  recovered without error  from a discrete
sample taken with a sampling rate  that is at least twice the maximum frequency
present in the signal (the Nyquist critical rate).    This
principle defines the choice of the sampling rate in almost all signal processing protocols.

It is known that, for signals with certain structure, the Nyquist rate could be excessive for signal recovery; see e.g. \cite{BS,ME}.  For example,  it is known that a sparse enough subsequence or an one-sided semi-infinite subsequence  can be removed from
an oversampling sequence \cite{F95,V87}.\index{(Vaidyanathan  (1987),  Ferreira (1995)).}

Some  paradigm changing results were obtained in \cite{CT06,CR1,CR,Don06} and consequent papers
in the so-called {\em "compressive sensing"} setting for finite sequences. Methods for
these sequences  can be immediately applied to digital processes and computer algorithms
since they would not require adaptation to inevitable data truncation, unlike results obtained  for continuous processes or for infinite discrete time processes.

The compressive sensing  explores sparsity of signals, i.e. restrictions on the number of non-zero members
of  the underlying finite sequences;  the location of these non-zero  members is not specified and is assumed to be unknown.
   The main result of
\cite{CT06,CR1,CR,Don05,Don06,Don11} was a new method of signal recovery  from a relatively small number
of measurement in frequency domain given certain sparsity in time domain.  (Equivalently, this result can be reformulated for measurements on time domain given the sparsity in frequency domain).   The recovery algorithm
suggested was based on $\ell_1$-minimization / Basis Pursuit Denoising method.

\def\SS{S}
 Quantification of recoverability criterions was presented in the form of an asymptotic estimate of the required number $|U|$ of observed Fourier coefficients  versus $\SS $, where $\SS $ is the
number of nonzero members of the underlying  finite sequences with $N$ elements. It was shown  \cite{CR1}  that recovery with overwhelming probability can be ensured given that $|U| \gtrsim C \SS \log (N)$  for some constant $C>0$.   There were some other modifications
such as  $\SS \lesssim |U|/\log(N/|U|)$  \cite{Don05}; see also \cite{Don06,Don11,HN,TG}. As  was mentioned in \cite{Don11}, these estimates
are not sharp and can be improved. Furthermore, asymptotically lossless linear recoverability for   $|D|\sim  \SS  + o(N)$   was proved  in \cite{WV} using Shannon-theoretic  setting under probabilistic assumptions  for i.i.d..
components of the underlying sequences with known distributions (which was essential).  It was shown therein that recoverability can be achieved with $|U|\sim  N\rho  + o(N)$, where  $\rho$ is the (upper) R\'enyi information dimension of the distribution.  Since $\rho\le \SS /N$ for sparse signals, it is a significant improvement. Moreover,  it was also shown also that this estimate for $|U|$ cannot be improved in this probabilistic setting \cite{WV}.  An impact of the noise contamination on compressed sampling was studied in \cite{BL,Chae}; various alternative setting were considered in \cite{C1,C2,C3,C4}. This illustrates how challenging is the problem of determination  of the allowed quantity of missed observations and the related problem of uniqueness of recovery result.

The present  paper considers data recovery problem for finite sequences and suggests a way to bypass  solution of this uniqueness problem given that a process  matching  available observations is found somehow.  This is achieved via imposing
  some restrictions on the process range described
  are defined by a special discretization of the spectrum range or the process range.
It appears that this approach allows to construct classes of sequences    that are
$\e$-dense  in the space of all sequences and, at the same time, have singleton uniqueness sets (Theorems \ref{ThM}, \ref{CSThM}-\ref{ZThM} below). The implied recovery procedure is neither  numerically feasible nor  stable since it would require to solve a Diophantine-type equation (equation
(\ref{eq}), (\ref{CSeq}), (\ref{Zeq}) below). However, any solution of the data recovery problem obtained by any method will be automatically a correct error-free solution. For example, the solution obtained via $\ell_1$-minimization
in the compressing sensing approach is guaranteed to be a correct one if all restrictions on matching the available observations are satisfied. To address  robustness of recoverability,  we considered  an alternative  setting where the data recovery is robust with respect to rounding of the input processes.  with rounded underlying processes  where observation of first $\SS $ Fourier coefficients is sufficient to recover sparse sequences with no more than $\SS $ nonzero terms (Theorem \ref{RThM}).

It can be noted that the  original ArXiv version of this paper included
Theorem  \ref{ThM} only; Theorem  \ref{CSThM} was added on 4 December  2017, Theorem \ref{RThM} was added on June 22 2018, Theorems \ref{CSTh} and \ref{ZThM} was added on June 22 2018.

\section{Some definitions and background}
For a integer $N>0$, let $\X$ be the  set of mappings $x:D\to\C^N$, where $D\defi\{0,1,...,N-1\}$.
This set can be associated with  the space  $\C^N$ as well as with the space of $N$-periodic sequences in $\C$.
We consider $\X$ as a linear normed space with the standard norm from $\C^N$.

 Let us consider the discrete Fourier transform  as a mapping  $\F:\X\to\X$ such that $\F(x)=Qx$, where
 $Q=\left\{\frac{1}{\sqrt{N}} e^{-i k t\pi/N } \right\}_{k,t=0}^{N-1}$  is  the DFT matrix,
$i=\sqrt{-1}$.

Let  $\nu$, $\nu_1$, $\mu$, and $\mu_1$,  be positive integers.

For $a\in \R$, $a\ge 0$, let $\lfloor a\rfloor= \{k\in\ZZ:\ a\in[k,k+1)\}$.
  For $a\in \R$, $a<0$, let $\lfloor a\rfloor= \{k\in\ZZ:\ x\in(k-1,k]\}$.
  For $a\in\R$, let  $\rho_{\nu,\mu}(a)= \nu^{-\mu}\lfloor \nu^{\mu}  a\rfloor$.
  We extend this function on complex numbers  such that
\baaa
\rho_{\nu,\mu}(z)=\rho_{\nu,\mu}(\Re z) +i\rho_{\nu,\mu}(\Im z), \quad z\in\C.
\eaaa
Similarly, we define rounding function $\rho_{\nu,\mu}:\C^N\to \C^N$, meaning the corresponding
component-wise rounding.

Let $\X_{\nu,\mu}= \rho_{\nu,\mu}(\X)$; this is the set of sequences from $\X$ with rounded components.

\section{Some cases where uniqueness sets are singletons}
\subsection{The case where  components of underlying process are observable}
In this section, we consider a problem of recovery of $y\in\X$ from available observations
of some of its components.

\begin{definition}\label{defU} Let  a subset $U$ of $D$  and a subset $\Y$ of  $\X$ be given.
If any $y\in\Y$ is uniquely defined by its trace $y|_U$, then we say that $U$ is an uniqueness set with respect to $\Y$.
\end{definition}
\begin{theorem}\label{ThM}
For any $\e>0$ and any $d\in \{1,...,N-1\}$, there exists  a  set $\Y_\e$
such that the following holds.
\begin{itemize}
\item[(i)]
The set $\Y_\e$ is closed in $\X$ and is such that if $\w y\in\Y_\e$ and $\ww y\in\Y_\e$ then $\w y-\ww y\in \Y_\e$.
\item[(ii)]
The set $\Y_\e$ is $\e$-dense in $\X$.
\item[(iii)]
The singleton  $U=\{d\}$ is a uniqueness set with respect to $\Y_\e$.
\end{itemize}
\end{theorem}

{\em Proof}.  Let $d\in\{1,...,N-1\}$ be fixed. To prove the theorem, we construct required sets $\Y_\e$ as sets of sequences with restrictions on their spectrum range.

Let $\xi=(\xi_0,...,\xi_{N-1})\in\X$ be defined such that
\baaa
\xi_k=e^{i 2(\rho_{\nu_1,\mu_1}(\pi)-\pi)dk/N}.\eaaa

 Let $\Y_{d,\nu,\mu,\nu_1,\mu_1}$ be the set of all $y\in\X$
 such that there exists
 $X=(X_0,...,X_{N-1})\in \X_{\nu,\mu}$ such that
 $Y_k=\xi_k X_k$ for $k=0,1,...,N-1$, where $Y=(Y_0,...,Y_{N-1})=\F y$.

It can be noted that a set $\Y_{d,\nu,\mu,\nu_1,\mu_1}$ is defined by restrictions on the range of the spectrum $Y=\F y$
of its members.

Clearly, for any $\e>0$, there exist large enough  $\nu$, $\nu_1$, $\mu$, and $\mu_1$,  such that
the set $\Y_{d,\nu,\mu,\nu_1,\mu_1}$ is $\e$-dense in $\X$.
In addition,  condition (i) in Theorem \ref{ThM} is satisfied for $\Y_{d,\nu,\mu,\nu_1,\mu_1}$ for all $\nu$, $\nu_1$, $\mu$, and $\mu_1$.

 Let $y\in\Y_{d,\nu,\mu,\nu_1,\mu_1}$, and let $Y=(Y_0,...,Y_{N-1})=\F y$, $X=(X_0,...,X_{N-1})\in \X_{\nu,\mu}$, $Y_k=\xi_k X_k$.
We have that
\baaa
y_d
=\frac{1}{\sqrt{N}}\sum_{k=0}^{N-1} e^{i 2\pi dk /N} Y_k=\frac{1}{\sqrt{N}}\sum_{k=0}^{N-1} e^{i 2\pi dk /N}\xi_k X_k  .
\label{eq1}\eaaa
Let \baaa
\o_k\defi  2\rho_{\nu_1,\mu_1}(\pi) dk/N.
\eaaa
By the choice of $\xi_k$ and by the definitions, we have that  $Y_0=X_0$ and
\baa y_d
=\frac{1}{\sqrt{N}}\left[X_0+\sum_{k=1}^{N-1} e^{i \o_k} X_k \right].
\label{eq}
\eaa

To prove the theorem, it suffices to show that, for any $\nu$, $\nu_1$, $\mu$,  $\mu_1$, and any $y\in\Y_{d,\nu,\mu,\nu_1,\mu_1}$, there exists an unique  $X=(X_0,...,X_{N-1})\in \X_{\nu,\mu}$
satisfying (\ref{eq}).

For this, it suffices to show that, for any $\nu$, $\nu_1$, $\mu$, and $\mu_1$,  we have that
 if $\w y_d=\ww y_d$ for some $\w y,\ww y\in\Y_{d,\nu,\mu,\nu_1,\mu_1}$, then  $\w y=\ww y$.

We are now in the position to complete the proof.
By condition (i) in Theorem \ref{ThM},  it suffices to show that if $ y_d=0$ for $y\in\Y_{d,\nu,\mu,\nu_1,\mu_1}$, then equation  (\ref{eq}) has only zero solution $X$ in $\X_{\nu,\mu}$. Let us show this.

Since $y\in\Y_{d,\nu,\mu,\nu_1,\mu_1}$, it follows from the definitions that  $X_k$ are rational numbers for $k=0,1,...,N-1$.
In addition,  $\o_k$  are rational numbers as well.  By the
Lindemann--Weierstrass Theorem, it follows that $X_k=0$ for all $k$
(see \cite{Baker}, Chapter 1, Theorem 1.4).
This completes the proof. $\Box$.
\begin{remark}\label{rem1} {\rm Theorem \ref{ThM} allows the following  obvious modification: for any $d\in D$,  $\e>0$, and any set $G\subset\X$, there exists
a $\e$-dense in $\X\cap G$ set $\Y_\e\subset \X\cap G$ such that  its uniqueness set is a singleton. }
\end{remark}
\subsection{The case where Fourier coefficients   are observable}
In this section, we consider a setting where, for a given  $y\in\X$, we observe some components of $Y=Q y$.

\begin{definition}\label{CSdefU} Let  a subset $U$ of $D$  and a subset $\Y$ of  $\X$ be given.
If any $y\in\Y$ is uniquely defined by the trace $Y|_U$, where  $Y=Q y$,
then we say that $U$ is a uniqueness set   in the frequency domain with respect to $\Y$.
\end{definition}

Let $\X_S$ be the set of all $y\in\X$ such that $\sum_{k\in D}\Ind_{\{y_k\neq 0\}}\le S$.
 Let $S\in \{1,...,N-1\}$ be given.
 \begin{theorem} \label{CSTh} 
\begin{enumerate}
\item If $N$ is a prime number, then any set $U\subset D$ such that $|U|= 2S$  is a uniqueness set   in the frequency domain with respect to $\Y_S$. (\cite{CR1}, Theorem 1.1).
\item
Let $U\subset D$ be such that $|U|=2S$ and that there exists $u\in D$ such that $U=\{u,u+1,...,u+2S\}$.
Then $U$ is is a uniqueness set   in the frequency domain with respect to $\Y_S$.
\end{enumerate}
\end{theorem}

\def\MM{{\rm M}}
{\em Proof of Theorem \ref{CSTh}  (ii)}. Let $\MM\defi |U|=2S$. 
Let $\TT\subset D$ be such that $|\TT|=\MM$, $\TT=\{t_1,...,t_{\MM}\}$.
Consider the matrix
 \baaa
 Q_{U,\TT}=\frac{1}{\sqrt{N}}\left(
   \begin{array}{cccc}
     e^{-i ut_1 \pi/N} &  e^{-i ut_2\pi/N}   & ... &  e^{-i ut_\MM\pi/N} \\
       e^{-i (u+1)t_1\pi/N} & e^{-i (u+1)t_2\pi/N}  & ... &  e^{-i (u+1)t_\MM\pi/N} \\
       ... & ... & ... & ... \\
       e^{-i (u+\MM)t_1\pi/N} &  e^{-i (u+\MM)t_2\pi/N} &  ... &  e^{-i (u+M)t_\MM\pi/N} \\
   \end{array}
 \right)\in\C^{\MM\times \MM}.
 \eaaa
 If $u=0$ then this is a non-degenerate Vandermonde  matrix with a nonzero  determinant that we denote $V$.
 If $u\neq 0$ then $|\det Q_{U,\TT}|=\left|V\prod_{t\in\TT} e^{-i u t \pi/N}\right|\neq 0$. The remaining part of the proof repeats the  proof of    Theorem 1.1  from \cite{CR1}. $\Box$
 
The following theorem  shows that the recovery uniqueness can be ensured for much smaller sets given additional restrictions on the processes range. 

\begin{theorem}\label{CSThM}
For any $\e>0$ and any $d\in \{1,...,N-1\}$, there exists  a  set $\w\Y_\e$
such that the following holds.
\begin{itemize}
\item[(i)]
The set $\w\Y_\e$ is closed in $\X$ and is such that if $\w y\in\w\Y_\e$ and $\ww y\in\w\Y_\e$ then $\w y-\ww y\in \w\Y_\e$.
\item[(ii)]
The set $\w\Y_\e$ is $\e$-dense in $\X$.
\item[(iii)]
The singleton  $U=\{d\}$ is a uniqueness set   in the frequency domain with respect to $\w\Y_\e$.
\end{itemize}
\end{theorem}

{\em Proof}. The proof is similar to the proof of Theorem \ref{ThM}; however, we provide it for the sake of completeness.

Let $d\in\{1,...,N-1\}$ be fixed. To prove the theorem, we construct required sets $\w\Y_\e$ as sets of sequences with restrictions on their range.

Let $\zeta=(\zeta_0,...,\zeta_{N-1})\in\X$ be defined such that
\baaa
\zeta_k=e^{-i 2(\rho_{\nu_1,\mu_1}(\pi)-\pi)dk/N}.\eaaa

 Let $\w\Y_{d,\nu,\mu,\nu_1,\mu_1}$ be the set of all $y\in\X$
 such that there exists $x=(x_0,x_1,...,x_{N-1})\in\X_{\nu,\mu}$ such that
 $y_k=\zeta_k x_k$ for $k=0,1,...,N-1$.

It can be noted that a set $\w\Y_{d,\nu,\mu,\nu_1,\mu_1}$ is defined by restrictions on the range
of its members $y$.

Clearly, for any $\e>0$, there exist large enough  $\nu$, $\nu_1$, $\mu$, and $\mu_1$,  such that
the set $\w\Y_{d,\nu,\mu,\nu_1,\mu_1}$ is $\e$-dense in $\X$.
In addition,  condition (i) in Theorem \ref{CSThM} is satisfied for $\w\Y_{d,\nu,\mu,\nu_1,\mu_1}$ for all $\nu$, $\nu_1$, $\mu$, and $\mu_1$.

 Let $y=(y_0,...,y_{N-1})\in\w\Y_{d,\nu,\mu,\nu_1,\mu_1}$,  let $Y=(Y_0,...,Y_{N-1})=\F y$, and let $x=(x_0,...,x_{N-1})\in \X_{\nu,\mu}$ be such that   $y_k=\zeta_k x_k$.
We have that
\baaa
Y_d
=\frac{1}{\sqrt{N}}\sum_{k=0}^{N-1} e^{-i 2\pi dk /N} y_k=\frac{1}{\sqrt{N}}\sum_{k=0}^{N-1} e^{-i 2\pi dk /N}\zeta_k x_k.
\label{CSeq1}\eaaa

By the choice of $\zeta_k$ and by the definitions, we have that  $y_0=x_0$ and
\baa Y_d
=\frac{1}{\sqrt{N}}\left[x_0+\sum_{k=1}^{N-1} e^{-i \o_k} x_k \right],
\label{CSeq}
\eaa
where \baaa
\o_k= 2\rho_{\nu_1,\mu_1}(\pi) dk/N.
\eaaa\par
To prove the theorem, it suffices to show that, for any $\nu$, $\nu_1$, $\mu$, and $\mu_1$, and any $y\in\w\Y_{d,\nu,\mu,\nu_1,\mu_1}$, there exists an unique  $x=(x_0,...,x_{N-1})\in \X_{\nu,\mu}$
satisfying (\ref{CSeq}).

For this, it suffices to show that, for any $\nu$ and $\mu$,  we have that
 if $\w Y_d=\ww Y_d$ for some $\w y,\ww y\in\X_{\mu,\nu}$,  $\w Y=\F (\w y)$, and $\ww y=\F(\ww y)$,
 then  $\w y=\ww y$.

We are now in the position to complete the proof.
By condition (i) in Theorem \ref{CSThM},  it suffices to show that if $ Y_d=0$ for $y\in\X_{\nu,\mu}$ and $Y=\F y$, then equation  (\ref{CSeq}) has only zero solution $x$ in $\X_{\nu,\mu}$. Let us show this.

Since $y\in\w\X_{\nu,\mu}$, it follows from the definitions that  $x_k$ are rational numbers for $k=0,1,...,N-1$.
In addition,  $\o_k$  are rational numbers as well.  By the
Lindemann--Weierstrass Theorem again, it follows that $x_k=0$ for all $k$
(see \cite{Baker}, Chapter 1, Theorem 1.4).
This completes the proof. $\Box$.
\begin{remark}\label{CSrem1} {\rm Similarly to Theorem \ref{ThM}, Theorem \ref{CSThM} allows the following  obvious modification: for any $d\in S$,  $\e>0$, and any set $\w G\subset\X$, there exists
a $\e$-dense in $\X\cap\w G$ set $\w\Y_\e\subset \X\cap\w G$ such that  its uniqueness set in the frequency domain is a singleton. }
\end{remark}
\subsection{The case where Z-transform is  observable}
In this section, we consider a setting where, for a given  $y\in\X$, we observe some values of its
Z-transform $Y=Z y$ defined as
\baaa
Y(z)=\sum_{k=0}^N z^{-k}y_k,\quad z\in \C.
\eaaa

Let $\T\defi \{z\in\C:\  |z|=1\}$.

\begin{definition}\label{ZdefU} Let  a subset $U$ of $\T$  and a subset $\Y$ of  $\X$ be given.
If any $y\in\Y$ is uniquely defined by the trace $Y|_U$, where  $Y=Z y$,
then we say that $U$ is a uniqueness set  in the frequency domain with respect to $\Y$.
\end{definition}
\begin{theorem}\label{ZThM}
For any $\e>0$ and any algebraic number $\o\in (-\pi,\pi]\setminus \{0\}$, there exists  a  set $\ww\Y_\e$
such that the following holds.
\begin{itemize}
\item[(i)]
The set $\ww\Y_\e$ is closed in $\X$ and is such that if $\ww y\in\ww\Y_\e$ and $\oo y\in\ww\Y_\e$ then $\ww y-\oo y\in \ww\Y_\e$.
\item[(ii)]
The set $\ww\Y_\e$ is $\e$-dense in $\X$.
\item[(iii)]
The singleton  $U=\{e^{i \o}\}$ is a uniqueness set   in the frequency domain with respect to $\ww\Y_\e$.
\end{itemize}
\end{theorem}

{\em Proof}. The proof is similar to the proof of Theorems \ref{ThM}-\ref{CSThM}; we provide it for the sake of completeness.

Let an algebraic number $\o\in(-\pi,\pi]\setminus \{0\}$ be fixed. To prove the theorem, we construct required sets $\ww\Y_\e$.

Clearly, condition (i) in Theorem \ref{ZThM} is satisfied for the sets $\X_{\nu,\mu}$ for all $\nu$ and $\mu$.
In addition,
for any $\e>0$, there exist large enough  $\nu$ and  $\mu$,  such that
the set $\X_{\nu,\mu}$ is $\e$-dense in $\X$. Let $\ww\Y_\e$ be selected as the
corresponding set    $\X_{\nu,\mu}$ selected for this $\e$.

 Let $y=(y_0,...,y_{N-1})\in\X_{\mu,\nu}$,  and let $Y=\Z y$. We have that
\baa
Y\ew
=\sum_{k=0}^{N-1} e^{-i \o k} y_k.
\label{Zeq}\eaa
\par
To prove the theorem, it suffices to show that, for any $\nu$ and $\mu$, and any
$y\in\X_{\nu,\mu}$, there exists at most one  $X=(X_0,...,X_{N-1})\in \X_{\nu,\mu}$
satisfying (\ref{Zeq}).

For this, it suffices to show that, for any $\nu$ and $\mu$,  we have that
 if $\ww Y\ew=\oo \Y\ew$ for some $\ww y,\oo y\in\X_{\nu,\mu}$,  $\ww Y=\Z \w y$, and $\oo Y=\Z\oo y$,
 then  $\ww y=\oo y$.

Furthermore, by condition (i) in Theorem \ref{ZThM},  it suffices to show that if $ Y\ew=0$ for $y\in\X_{\nu,\mu}$ and $Y=\Z y$, then equation  (\ref{Zeq}) has only zero solution $x$ in $\X_{\nu,\mu}$. Let us show this.

Since $y\in\X_{\nu,\mu}$, it follows that the components of $y$ are rational numbers.
In addition,  $i\o k$  are algebraic  numbers.  By the
Lindemann--Weierstrass Theorem again, it follows that $x_k=0$ for all $k$
(see \cite{Baker}, Chapter 1, Theorem 1.4).
This completes the proof. $\Box$.
\begin{remark}\label{Zrem1} {\rm Similarly to Theorems \ref{ThM}-\ref{CSThM},  Theorem \ref{ZThM} allows the following  obvious modification: for any algebraic number
 $\o\in(-\pi,\pi]\setminus \{0\}$, any $\e>0$, and any set $\ww G\subset\X$, there exists
a $\e$-dense in $\X\cap\ww G$ set $\ww\Y_\e\subset \X\cap\ww G$ such that  its uniqueness set in the frequency domain is a singleton. }
\end{remark}

\section{The case of multiple observations}
\subsection{Extended systems with additional observations}
Assume that, in the setting of Theorem  \ref{ThM}, there are available observations of $y_t$ for  $ y=(y_0,...,y_{N-1})\in\Y_{d,\nu,\mu,\nu_1,\mu_1}$  at
$t\in D_1\cup\{d\}$, where $D_1$ is a subset of $D$.  In this case,  equation (\ref{eq}) can be supplemented
with equations
  \baaa
&&y_t
=\frac{1}{\sqrt{N}}\Bigl[Y_0+\sum_{k=1}^{N-1} e^{i 2\pi t k/N}\xi_k X_k \Bigr],\quad t\in D_1.\hphantom{xxxx}
\label{eqm}
\eaaa
This system has a unique solution, since even a single equation with $t=d$ has a unique solution. Therefore, any solution of this system (for example, obtained via minimization of
$\|Y\|_{\ell_1}$ as in the compressive sensing approach) ensures an error-free recovery of the underlying process.

Similar reasoning can be applied in the setting of Theorem  \ref{CSThM}:  that, in the setting of the proof of this theorem,
there are available observations of   $Y_\o$ for  $y\in\w\Y_{d,\nu,\mu,\nu_1,\mu_1}$ and $Y=(Y_0,...,Y_{N-1})=\F y$  at
$\o\in D$, where $D$ is a subset of $S$ such that $d\in D$.  In this case,  equation (\ref{CSeq}) can be replaced by a system of  equations
  \baaa
&&Y_\o
=\frac{1}{\sqrt{N}}\Bigl[y_0+\sum_{k=1}^{N-1} e^{-i 2\pi \o k/N} y_k \Bigr],\quad \o\in D\hphantom{xxxx}
\label{CSeqm}
\eaaa
for an unknown vector $y=\{y_k\}\in\X$ such that $\y_k =\zeta_k x_k$.
We know that this system has a unique solution, since even a single equation with $\o=d$ has a unique solution. Therefore, any solution of this system (for example, obtained via minimization of
$\|y\|_{\ell_1}$ as in the compressive sensing approach) ensures an error-free recovery of the underlying process.


If the sets  $\w G$ in Remarks  \ref{rem1} and \ref{CSrem1} are bounded, then the  sets  $\w\Y_{d,\nu,\mu,\nu_1,\mu_1}\cap\w G$ and $\w\Y_{d,\nu,\mu,\nu_1,\mu_1}\cap\w G$ are finite. In this case,  for certain range of $N$, $G$, and $\w G$,  the solution of equations  \index{(\ref{eq})} and (\ref{CSeqm}) can be obtained with a brute-force search.  Ever-growing available computational power will allow  larger and larger $N$, $G$, and $\w G$.

Similar reasoning can be applied in the setting of Theorem  \ref{ZThM}.

\subsection{Robustness with respect to rounding for sparse signals with additional observations}
In this section, we consider a setting where, for a given  $x\in\X$, we observe some components of $X=Q x$.
\def\w{\widehat}

\begin{definition}\label{defR} Let  a subset $U$ of $D$ of cardinality $|U|$
 and a subset $\Y$ of  $\X$ be given such that
 $U$ is a uniqueness set   in the frequency domain with respect to $\Y$ in teh sense of Definition \ref{CSdefU}.
 Let $A: \C^{|U|}\to\C^N$ be a mapping such that $A(X|_U)=x$, where $X= Q x$, i.e. this mapping represents a recovery algorithm of $x$ from $X|_U$. We  say that this algorithm is robust with respect to data rounding if,
for any $\d>0$,  there exists $\oo\mu=\oo\mu(\d,N,U)>0$ such that
 $|\w x_{\nu,\mu}-x|\le\d$ for any $\mu\ge \oo\mu$ and any $x\in\Y$ such that $|x|\le 1$.
Here $\w x_{\nu,\mu}=A(X_{\nu,\mu}|_U)$, where $X_{\nu,\mu}=Q(R_{\nu,\mu}(x))$.
\end{definition}
In the definition above,   the estimate $\w x_{\nu,\mu}$ of $x$ is obtained as the output of
the corresponding algorithm with the rounded input process.

 For an integer $S\in\{1,...,N\}$, let $\X_{S}$ be set of all
$x\in \X$ with no more than  $S$ non-zero components.

\begin{theorem}\label{RThM}
For any $\e>0$, there exists  a  set $\Y_\e$
such that the following holds.
\begin{itemize}
\item[(i)]
The set $\Y_\e$ is $\e$-dense in $\X$.
\item[(ii)]  The  set $U= \{0,1,...,S-1\}$  is a uniqueness set   with respect to $\Y_\e\cap \X_S$.
\item[(iii)] There exists an algorithms of recovery  $x\in\Y_\e\cap \X_S$  from $X|_U$ that is robust with respect to data rounding in the  sense of Definition \ref{defR}.
\end{itemize}
\end{theorem}

{\em Proof of Theorem \ref{RThM}}.
For $a\in \R$, $k\in\ZZ$, let function $p_{\nu,k}(a):\R\to \{0,1,.., \nu\}$ be defined as the corresponding term
in the representation
\baaa
a=\sum _{k=-\infty}^\infty p_{\nu,k}(a)\nu^{-k}.
\eaaa
We extend this function on complex numbers  such that
\baaa
p_{\nu,k}(z)=p_{\nu,k}(\Re z) +ip_{\nu,k}(\Im z), \quad z\in\C.
\eaaa
\index{Similarly, we define  function $p_{\nu,k}:\C^N\to \C^N$, meaning corresponding
component-wise application of the corresponding functions.}

For $z\in\C$ and integers $k\in\{0,1,...,N-1\}$,
$M>0$, and $\nu>0$, let $\zeta_{\nu,M,k}(z)\in\C$ be defined such that
\baa
\zeta_{\nu,M,k}(z)=\rho_{\nu,M}(z )+\nu^{-M-k}\Ind_{z\neq 0}.
\label{zeta}\eaa
By the definition, \baaa
&&R_{\nu,M}(\zeta_{\nu,M,k}(z))=R_{\nu,M}(z),\quad
\breakk  p_{\nu,M+k}(\zeta_{\nu,M,k}(z))=\Ind_{z\neq 0},\\ &&
   p_{\nu,M+m}(\zeta_{\nu,M,k}(z))=0,\quad m\in\ZZ,\quad m>0,\quad m\neq k.
\eaaa
 Let $\Y_{\nu,M}$ be the set of all $x=(x_0,x_1,...,x_{N-1})\in\X$
 such that   $x_k=\zeta_{\nu,M,k}(x_k)$ for $k=0,1,...,N-1$.

It can be noted that the class $\Y_{\nu,M}$ is defined by restrictions of the rounding type on the range
of its members.

Let us show that, for any integer $\nu\ge 2$, the sets  $\Y_{\nu,M}$ are such as required in the theorem statement.

Clearly, for any $\e>0$ and $\nu\ge 2$, there exist large enough  $M$,  such that
the set $\Y_{\nu,M}$ is $\e$-dense in $\X$. Hence the required property (i) holds.

Let $\w\Y_{\nu,M,S}$ we the set of all vectors $(X_0,...,X_S)$ such that there exists
$x=(x_0,...,x_{N-1})\in \Y_{\nu,M}\cap \X_S$ such that
 \baa
&&X_\o
=\frac{1}{\sqrt{N}}\sum_{k=0}^{N-1} e^{-i 2\pi \o k/N} x_k,\quad \o\in U, \quad
 \breakk\hbox{subject to}\quad  x\in \Y_{\nu,M}\cap \X_S.\hphantom{xxxx}
\label{Reqm}
\eaa

In particular,
 \baa
&&X_0
=\frac{1}{\sqrt{N}}\sum_{k=0}^{N-1} x_k.\hphantom{xxxx}
\label{Reqm0}
\eaa
By the definitions, it follows
 that \baa
p_{\nu,M+k}(X_0)=\Ind_{\{x_k\neq 0\}}.\label{px}
\eaa

Let us  show that, for any $\nu$ and $M$, and any
$(X_0,...,X_S) \in\w\Y_{\nu,M,S}$,
system (\ref{Reqm})  has  a unique solution  $x\in \Y_{\nu,M,S}\cap\X_S$.

\par
Let $K(x)=\{k_1,....,k_S\}\subset D$ be  a set  such   that  $\supp x\subset K(x)$.
For certainty, we presume that this set is formed from minimal possible numbers.
Since  $p_{\nu,M+k}(X_0)=0$ for $k\notin K(x)$,
it follows that the sets of solution for system  (\ref{Reqm}) is the same as for the system \baa&& X_\o
=\frac{1}{\sqrt{N}}\sum_{k\in K(x)} e^{-i 2\pi \o k/N} x_k,\quad \o\in U, \quad
 \breakk \hbox{subject to}\quad x\in \Y_{\nu,M}\cap \X_S.\hphantom{xxxx}
\label{SK}
\eaa
Let $Q_{K(x)}\in \C^{S\times S}$ be the matrix of this system. Similarly to the proof of Theorem \ref{CSTh}(ii), where we  use now $\MM=S$ and $u=0$, we obtain that this is a  Vandermonde matrix. Hence  the system has a unique solution which  is also an exact solution  of the problem of recovery  of $x$ from
observations $(X_0,X_1,...,X_{S-1})$. Hence the required property (ii) holds.

Furthermore, let us prove that the property  (iii) holds for the recovery algorithm
consisting of calculation of $K(x)$ and consequent solution of system (\ref{SK}) as described in the proof above.

Let us observe first that $K(x)=K(R_{\nu,\oo\mu}(x))$ for
$\oo\mu=M+2^N$  for any $M>0$ and $x\in \X_{\nu,M}\cap \X_S$.

Furthermore,
 $\sup_{x\in\X_\S}\|Q_{K(x)}^{-1}\|<+\infty$ since there exists a
finite number of possible choices for $K(\cdot)$. Here $\|\cdot\|$ is the Frobenius matrix norm.

Clearly, one can select large enough integer $\mu>0$ such that
\baaa
|R_{\nu,\mu}(x)- x| \le\d \sup_{x\in\X_S} \|Q_{K(x)}^{-1}\|\quad \forall x\in \X: \quad |x|\le 1.
\eaaa
If, in addition,  $\mu\ge \oo\mu$, then $K(R_{\nu,\mu}(x))=K(x)$ and   \baaa
|Q_{K(x)}^{-1}R_{\nu,\mu}(x)- Q_{K(x)}^{-1}x| \le \d.
\eaaa
Hence the required property (iii) holds.
This completes the proof of Theorem \ref{RThM}. $\Box$.

\begin{remark}
If $N$ is  a prime number, then, by Chebotarev Lemma (see, e.g. \cite{Cheb}), the matrix $Q_{K(x)}$ of system (\ref{SK}) is non-degenerate  for any choice of  $U=\{\o_1,....,\o_S\}\subset D$.  In this case, Theorem \ref{RThM}  can be extended on the case of  $U=\{\o_1,....,\o_S\}\subset D$ such that $\o_1=0$ and with arbitrarily selected $\{\o_j\}_{j>1}$.
\end{remark}
\begin{remark} As can be seen form the proof, recovery of the set $K(x)$   requires quite precise
representation of $X_0=\sum_{k=0}^{N-1}x_k$, which is numerically challenging for large $N$. The robustness of recovery established in the theorem takes effect for quite large $\mu$ only.
\end{remark}

\section{Discussion and future research}\label{secCon}
Traditionally, possibilities of data recovery and extrapolation are associated with spectrum degeneracy such as bandlimitiness, the presence of spectrum gaps, and data sparsity.
Theorems \ref{ThM}, \ref{CSThM}-\ref{ZThM}  suggest to explore restraints on the process range or process spectrum. These theorems
establish that there are $\e$-dense sets of sequences that are uniquely defined
by  a single measurement.   The corresponding ranges are defined by a special type of discretization that involves  adjustment using $\xi_k$ or $\zeta_k$.  Sparsity,  bandlimiteness, or presence
spectrum gaps, are not required for this.

Theorems \ref{ThM}-\ref{ZThM}   do not lead to an efficient numerical algorithm. Formally, these theorems  and their proof imply a data compression and consequent recovery procedure.
For example, Theorem  \ref{CSThM} implies the following procedure: (i) a sequence $x\in\X$
can be approximated by some close enough   $y\in\w\Y_{d,\nu,\mu,\nu_1,\mu_1}$; (ii) this $y$
can be recovered via rational solutions  $\{x_k\}$ of equations \index{(\ref{eq}) or} (\ref{CSeq}) respectively   which are versions of Diophantine  equation.
 Currently, it is unclear what kind of computational power would be sufficient to solve these equations  what  quantity of information is required to code a "rounded" version  $x_d$  for a given class $\Y_\e$.  This problem is beyond  the scope of this paper; review of some related  methods and some references  can be found, e.g.,  in \cite{Chon,Smart}.

It appears that some  robustness with respect to rounding can be achieved in a setting with  sequences with rounded components under  additional restrictions on their sparsity and with additional observations of Fourier coefficients
(Theorem \ref{RThM}). In this setting, a different kind of rounding was used, comparing with Theorems \ref{ThM}-\ref{ZThM}. However, the recovery would require precise summation that  could be computationally expensive for large $N$.

\end{document}